\documentclass[a4paper]{article}

\usepackage{INTERSPEECH2021}
\usepackage{color,cite}
\usepackage{graphicx}
\usepackage{multirow,url}
\usepackage[labelformat=simple]{subcaption}

\title{Libri-adhoc40:\\ A dataset collected from synchronized ad-hoc microphone arrays}
\name{Shanzheng Guan, Shupei Liu, Junqi Chen, Wenbo Zhu, Shengqiang Li, Xu Tan, Ziye Yang,\\ Menglong Xu, Yijiang Chen, Jianyu Wang, Xiao-Lei Zhang}
\address{
  CIAIC, School of Marine Science and Technology, Northwestern Polytechnical University, China}
\email{gshanzheng@mail.nwpu.edu.cn, xiaolei.zhang@nwpu.edu.cn}

\begin{document}

\maketitle
\begin{abstract}
  Recently, there is a research trend on ad-hoc microphone arrays. However, most research was conducted on simulated data. Although some {datasets} were collected with a small number of distributed devices, they were not synchronized which hinders the fundamental theoretical research on ad-hoc microphone arrays. To address this issue, this paper presents a synchronized speech corpus, named \textit{Libri-adhoc40}, which collects the replayed Librispeech data from loudspeakers by ad-hoc microphone arrays of 40 strongly synchronized distributed nodes {in a real office environment. Besides, to provide the evaluation target for speech frontend processing and other applications, we also recorded the replayed speech in an anechoic chamber.} We trained several multi-device speech recognition systems on both the {Libri-adhoc40} dataset and a simulated dataset. Experimental results demonstrate the validness of the proposed corpus which can be used as a benchmark to reflect the trend and difference of the models with different ad-hoc microphone arrays. The dataset is online available at https://github.com/ISmallFish/Libri-adhoc40.
\end{abstract}
\noindent\textbf{Index Terms}: Libri-adhoc40, ad-hoc microphone arrays, distributed microphone arrays

\section{Introduction}

Deep learning based speech processing has made significant progress. However, the progress was mostly made with single-channel front-ends or multichannel front-ends on single devices \cite{wang2018supervised}. As we know, the performance of speech processing degrades significantly when the distance between the speech source and the microphone array receiver increases, which is known as the far-field speech processing problem. Fortunately, ad-hoc microphone array can significantly reduce the occurrence probability of far-field environments \cite{ZHANG2021101201}. It is a set of distributed microphones collaborating with each other \cite{raykar2004position}. Conventional methods try to organize the microphones in a blind way, which faces many challenges.

Recently, deep learning has been introduced to the study of ad-hoc microphone arrays \cite{ZHANG2021101201,yang2020deep,9053455,luo2019fasnet,wang2020neural,8932598,qin2020hi}, which provides a promising solution to the challenges. In \cite{ZHANG2021101201,yang2020deep}, a supervised channel selection strategy based on deep learning was proposed to group the distributed microphones with high signal-to-noise ratios (SNR) into a local microphone array. However, the aforementioned studies were conducted on simulated data only.

Some work was conducted on real-world data. In \cite{luo2019fasnet}, the authors first conducted single-channel speech separation on a selected reference microphone, and then estimated a beamforming filter for all remaining microphones based on the output of the reference microphone. In \cite{wang2020neural}, a novel neural network architecture was proposed to capture both the inter-channel and temporal correlations from the multi-channel input of ad-hoc microphone arrays. \cite{8932598} designed a speech recognition system which first makes all channels share the same encoder and then fuses all channels via stream attention.
\cite{qin2020hi} proposed a speaker recognition system based on ad-hoc microphone arrays. It first trains a single-channel speaker recognition system, then applies it to each channel, and finally fuses the outputs of the channels for the final decision. However, their experimental data was recorded with few devices only.

There are already some corpora collected with ad-hoc microphone arrays \cite{qin2020hi,barker2018fifth,watanabe2020chime6,7404805,corey2019massive}.
However, most of them were collected with a small number of distributed devices too. In \cite{qin2020hi}, a speaker verification dataset called HI-MIA was collected with 7 recording devices for both training and test. In \cite{barker2018fifth}, the CHiME-5 dataset employed 6 Kinect microphone arrays and 4 binaural microphone pairs to record natural
conversational speech. To our knowledge, the ad-hoc microphone array in the Massive Distributed Microphone Array dataset \cite{corey2019massive}, which consists of 4 wearable arrays and 12 tabletop arrays, is the largest array that has {ever been} used for recording publicly available data. As we known, when the ad-hoc nodes are too few, it is difficult to fully explore the potential of ad-hoc microphone arrays.

Moreover, none of the above corpora were collected with synchronized devices. Because the hardware and software processing pipelines between devices are different, the collected data may have significant variations \cite{9053074,8732945}. In \cite{watanabe2020chime6}, CHiME-6 synchronized the ad-hoc recordings of CHiME-5 via frame-dropping and clock-drift compensation. However, the synchronization technique misses the signal propagation delay information between devices.

To fascinate the fundamental research of ad-hoc microphone arrays on how well it can improve the performance ideally, we need to collect {synchronized data} from {large} ad-hoc microphone arrays, leaving the device synchronization problem as a separate topic at the current research stage. To address this issue, in this paper, we create a dataset, named \textit{Libri-adhoc40}, which collects the replayed Librispeech data \cite{panayotov2015librispeech} from loudspeakers by ad-hoc microphone arrays of 40 synchronized distributed microphones, where the `train-clean-100', `dev-clean' and `test-clean' subsets of Librispeech were used as the speech source. To provide the evaluation target for speech frontend processing and other applications, we also recorded the replayed speech in an anechoic chamber. Eventually, Libri-adhoc40 contains 4510 hours data in total with 110 hours data per microphone. We conducted a speech recognition evaluation on the test set of Libri-adhoc40, where both the simulated data and the training set of Libri-adhoc40 were used for model training. Experimental results demonstrate the validness of Libri-adhoc40.

{The rest of this paper is organized as follows.} We first overview the dataset and its recording method in Sections 2 and 3 respectively, then conduct a baseline evaluation in Section 4, and finally conclude in Section 5.

\begin{figure*}[t]
\hspace{4mm}
\begin{subfigure}{6.5cm}
\centering
	\includegraphics[width=7.1cm]{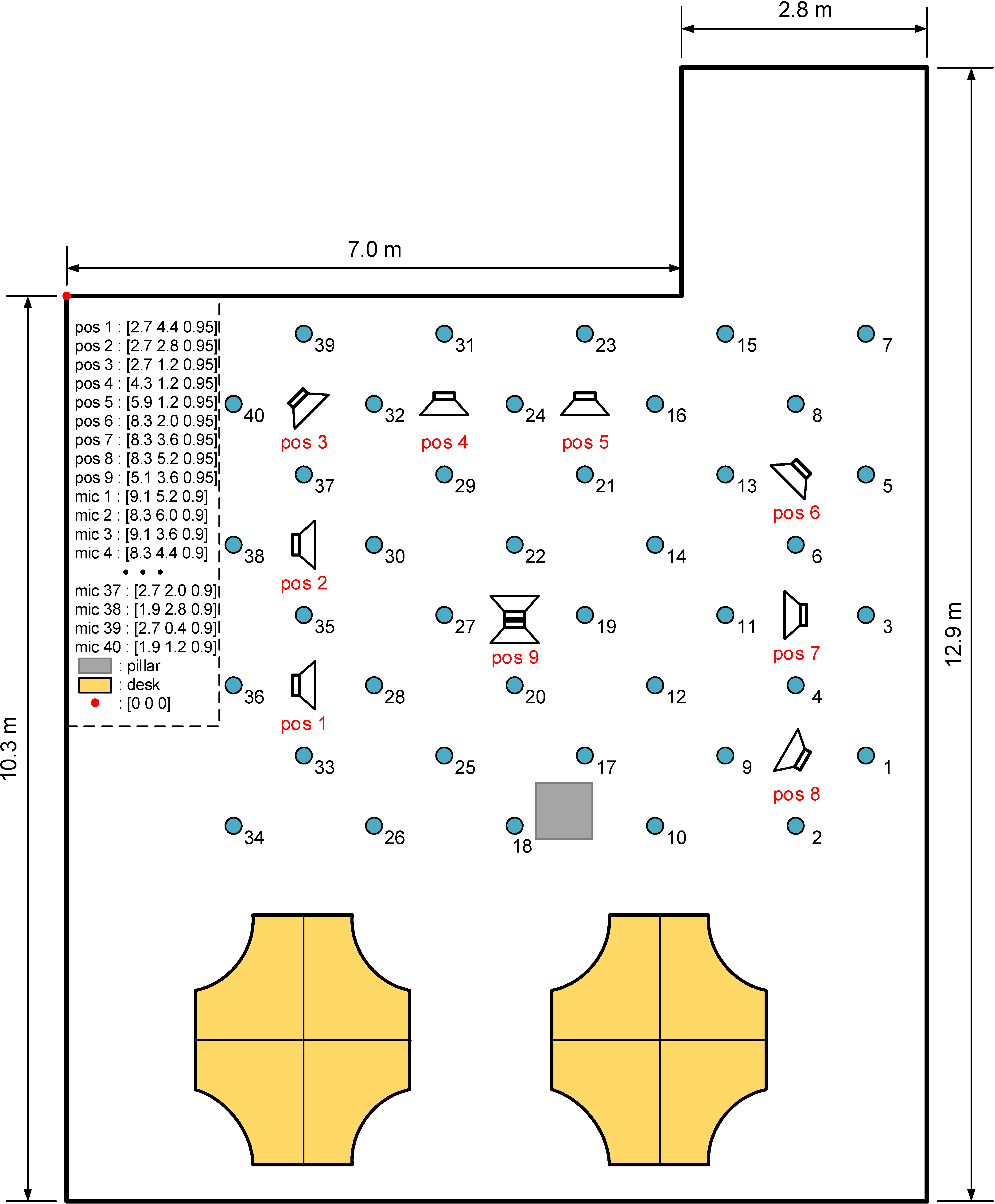}
	\caption{Training set}
	\label{fig:attention_a}
\end{subfigure}\hspace{24mm}
\begin{subfigure}{6.5cm}
\centering
	\includegraphics[width=7.1cm]{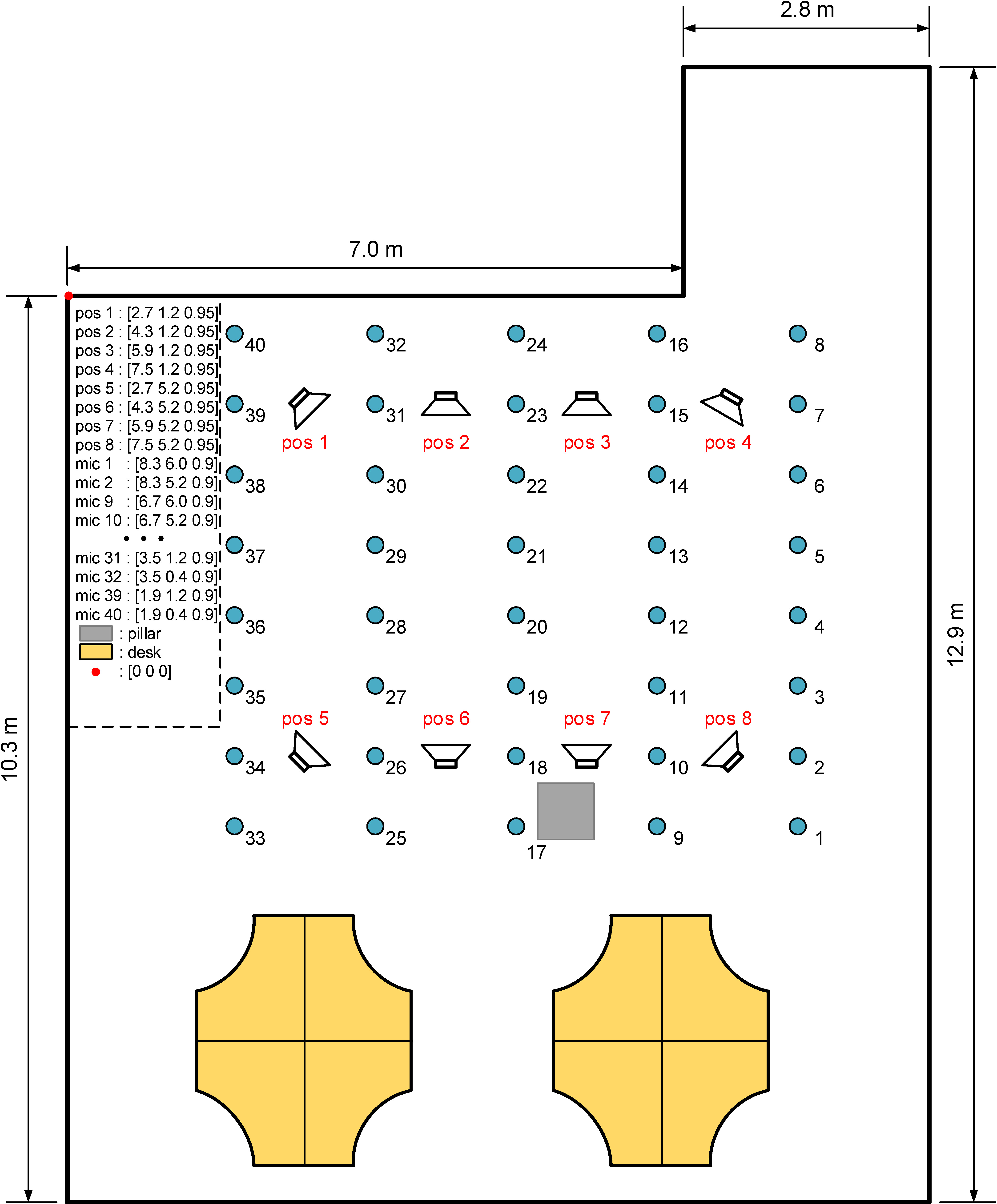}
	\caption{Development and test sets}
	\label{fig:attention_a}
\end{subfigure}
\caption{Recording environment and setting of Libri-adhoc40. The red dot indicates the origin of the reference axes.
The blue dots indicate the positions of the microphones, whose coordinates are listed in the upper-left corner. The positions and orientations of the loudspeaker are marked by loudspeaker icons. The terms `pos' is short for position. The term `mic' is short for microphone.}
\label{fig:rec_setting}
\end{figure*}

\section{Description of Libri-adhoc40}


The Librispeech corpus \cite{panayotov2015librispeech} {is} derived from audiobooks that are part of the LibriVox project. It contains 1000 hours of clean speech with a sampling rate of 16 kHz. The gender and per-speaker duration are reasonably balanced.
The Libri-adhoc40 dataset takes the `train-clean-100', `dev-clean', and `test-clean' subsets of Librispeech as the clean speech source, which contains about 110 hours of US English speech from 331 speakers.

\subsection{Recording environment} 
We replayed the subsets of Librispeech in an office room and an anechoic chamber individually which are described as follows:
\begin{itemize}
\item \emph{Office room}: The {plane structure} of the office room is shown in Figure~\ref{fig:rec_setting}. The height of the room is 4.2 m. Because the room size is large, and because the floor is laid with smooth tiles, the room is highly reverberant with the $T_{60}$ around 900 ms. Because the room is far from noisy environments, the recorded speech has little additive noise. A {directional} loudspeaker and 40 {omnidirectional} microphones of the same type were placed in the room. The sampling rate is 16 kHz.
\item \emph{Anechoic chamber}: The size of the net space of the anechoic chamber is $11.8{\times}4.2{\times}3.8$ m after the installation of sound-absorbing materials. The same loudspeaker and a handy recorder {were} placed in the anechoic chamber. The speech was recorded at 48 kHz, and further downsampled to 16 kHz.
\end{itemize}

\subsection{Training data}

As shown in Figure~\ref{fig:rec_setting}(a), the loudspeaker was placed at 9 positions with 10 orientations, where the loudspeaker at `pos 9' has 2 opposite orientations. The distances between the loudspeaker and the microphones are ranged from 0.8 m to 7.4 m. The speech source is the `train-clean-100' corpus of Librispeech, which contains 251 speakers. We replayed the corpus with about 20 to 40 speakers per position. A detailed configuration, including the coordinates of the loudspeaker and microphones, as well as the relationship between the speaker identities and the positions, are described at https://github.com/ISmallFish/Libri-adhoc40.

\subsection{Development and test data}

As shown in Figure~\ref{fig:rec_setting}(b), the loudspeaker was placed at 8 positions. The distances between the loudspeaker and the microphones are ranged from 0.8 m to 7.4 m as well. The positions of the loudspeaker and 40 microphones for preparing the development and test data are different from those for preparing the training data, which is designed for evaluating the generalization ability of speech processing algorithms on different array patterns. The speech sources for development and test are the `dev-clean' and `test-clean' corpora of Librispeech respectively, each of which contains 40 speakers. We replayed the corpus with 10 speakers per position.

\subsection{Ground-truth clean speech}

Because the loudspeaker may introduce {an} unwanted mismatch between the original recordings and the output of the loudspeaker, we replayed the clean speech of Librispeech  in the anechoic chamber to provide the ground-truth clean speech of Libri-adhoc40. The distance between the loudspeaker and the recording device is 40 cm. The {sound} volume of the loudspeaker was set the same as that in the office room.

\section{Methodology}

\subsection{Recording equipment}

The equipment for recording the data is listed in Table~\ref{tab:equip_info}. `JBL One Series 104' was selected as the loudspeaker. In this loudspeaker, a high-frequency driver aligned with a precisely contoured woofer cone is used to deliver accurate response. Because its treble and bass units are tightly arranged together, this design makes the loudspeaker behave like a point source.

\begin{table}[t]
  \caption{Recording equipment.}
  \label{tab:equip_info}
  \centering
\begin{tabular}{l|l|c}
\toprule
\multicolumn{1}{c|}{{\textbf{Device}}} & \multicolumn{1}{c|}{{\textbf{Product model}}} & {\textbf{Quantity}} \\ \hline
Microphone                  & Superlux ECM 999                        & 40   \\ \hline
Preamplifier                     & Focusrite Scarlett Octopre 8            & 4    \\ \hline
Sound card                  & RME Fireface UFX II                     & 2    \\ \hline
Handy recorder              & Zoom H1N                                & 1    \\ \hline
Loudspeaker                 & JBL One Series 104                                     & 2    \\ \bottomrule
\end{tabular}

\end{table}

To reduce the difference of the distortion between devices and eliminate device asynchronization, 40 `Superlux ECM 999' microphones and two `RME Fireface UFX II' sound cards were pre-synchronized.

Specifically, each sound card was connected with 20 microphones separately, where 4 microphones were connected to the sound card directly, and the other 16 microphones were connected indirectly through 2 `Focusrite Scarlett Octopre' microphone preamplifiers, each of which connects 8 microphones. After a careful evaluation, we find that the time delay caused by the preamplifiers can be neglected. To reduce the difference in gains between the two sets of channels, the gains of the sound cards and preamplifiers were adjusted in advance. A `Zoom H1N' handy recorder was employed to record anechoic signals at a sampling rate of 48 kHz. The recordings were further downsampled to 16 kHz manually.

\subsection{Recording process}

\begin{figure}[t]
  \centering
  \includegraphics[width=\linewidth]{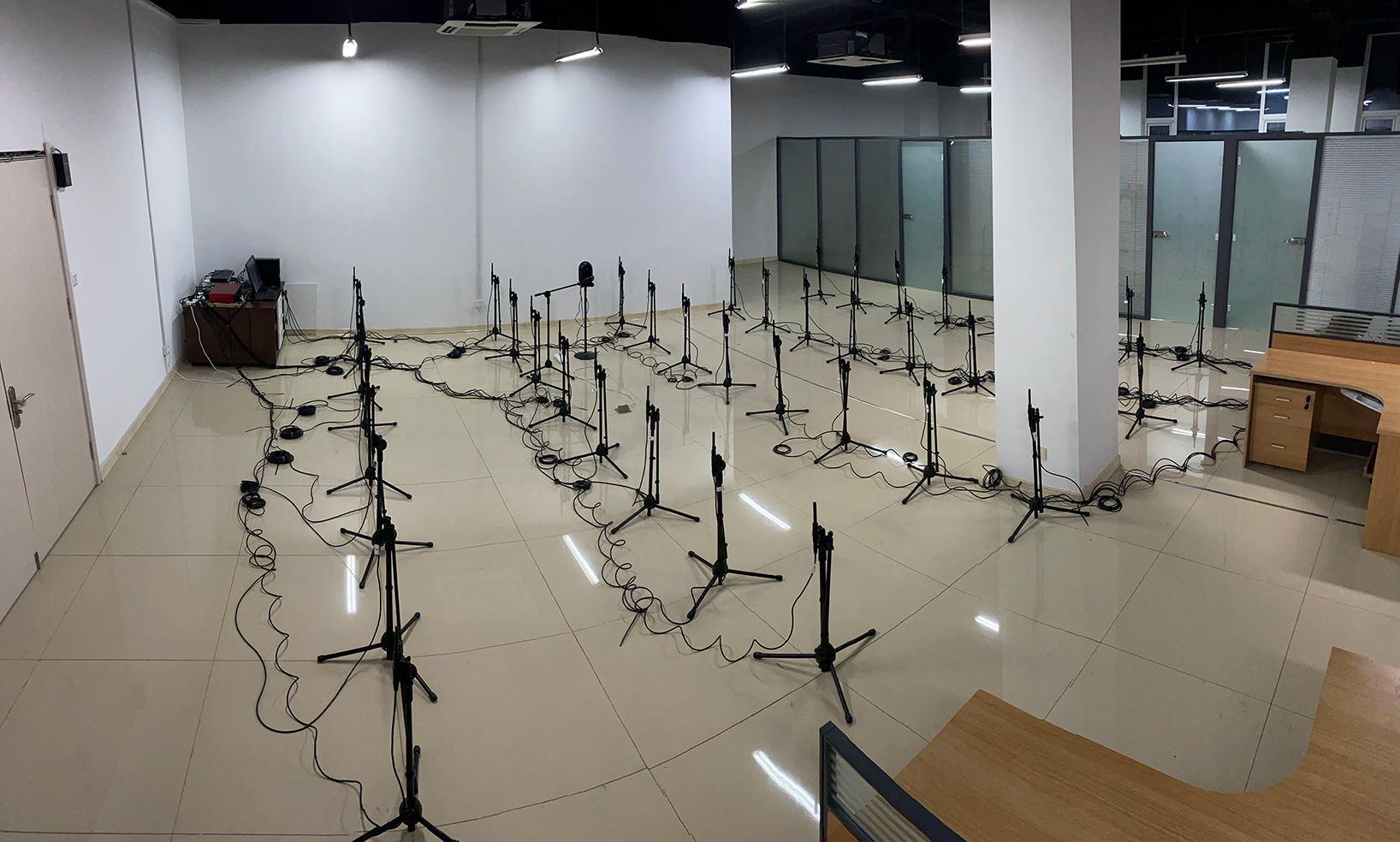}
  \caption{A picture of the recording environment for replaying the `dev-clean' and `test-clean' corpora. The loudspeaker was placed at \emph{pos 2}. The microphones 1 to 20 were connected to one sound card, while the microphones 21 to 40 were connected to the other sound card.}
  \label{fig:rec_envs}
\end{figure}

We played back Librispeech in a streaming fashion, where the sentences from the same speaker were concatenated into a sequence and played back continuously.
A picture of the real recording environment is shown in Figure~\ref{fig:rec_envs}.

\subsection{Postprocessing}

The two independent sound cards introduce a device asynchronization problem into the two sets of the microphones. It was mainly caused by (i) the asynchronization of the recording start time and (ii) the random drop of the sample points. To compensate the start time difference, we conducted a time delay estimation by playing white noise before the recording, which makes us possible in inferring the time delay difference.

Although the sample drop happened occasionally, the accumulation of the negative effect cannot be neglected if we play a long sequence continuously.
To compensate the sample drop caused by the two independent sound cards, we first carefully selected one microphone per sound card and then calculate the time difference of arrival between the two microphones for each position of the loudspeaker, before the data recording. Finally, if the time delay difference of the recorded data changed at some point-in-time, we compensated the detected sample drop at the time.

At last, we partitioned the recorded continuous speech according to the original segmentation lengths of the Librispeech utterances. Each partitioned segment was saved with the same name of its corresponding Librispeech utterance in a subdirectory that has the same name with Librispeech.

This strict synchronization setting makes the dataset generalizable for simulating device asynchronized situations by, e.g., performing bandpass filtering, waveform amplitude clipping, and delay perturbation operations \cite{wang2021continuous} to the data.

\section{Experiments}

In this section, we evaluate the validness of Libri-adhoc40 in an automatic speech recognition (ASR) task with ad-hoc microphone arrays.

\subsection{Datasets}

To evaluate the performance of the ASR with ad-hoc microphone arrays, we simulated a similar dataset with Libri-adhoc40, named \textit{Libri-adhoc40-simu}. Because all ASR systems in evaluation were tested on the test set of Libri-adhoc40, Libri-adhoc40-simu consists of only a training set and a development set, which were generated from the `train-clean-100' and `dev-clean' corpora of Librispeech respectively. The simulation environment is described as follows.

To roughly match the recording environments of Libri-adhoc40, we simulated a room with a size of $10{\times}10{\times}4$ m. Forty simulated microphones for both training and development were placed at the same locations as Libri-adhoc40. For each utterance, a simulated loudspeaker for playing back the utterance was placed \textit{randomly} in the room with its position located in the covering range of the ad-hoc microphone arrays and at least 0.6 meter away from the microphones; the room impulse response was generated by an image source model \cite{allen1979image}, where the $T_{60}$ was sampled from a
Gaussian distribution with a mean value of 0.7 second, a standard deviation of 0.1 second, a lower bound of 0.5 second, and an upper bound of 1.2 second.

We constructed three test scenarios. The first two scenarios randomly select 10 and 25 channels respectively for each test utterance. The third scenario uses all 40 channels for evaluation.

\subsection{ASR systems}

We used a single-channel conformer based automatic speech recognition (ASR) system \cite{gulati2020conformer} and a multichannel ASR based on the Scaling Sparsemax stream attention \cite{chen2021scaling} as the ASR systems, which are described as follows:

\textbf{Single-channel conformer (oracle one-best)}: We trained the single-channel ASR with the clean speech of the original Librispeech corpora directly. In the test stage, we picked the channel that was physically closest to and also faced by the loudspeaker as the input of the single channel ASR system.

\textbf{Scaling Sparsemax stream attention based multichanel ASR (Scaling Sparsemax)}: We first used the single-channel ASR trained with the clean speech of Librispeech as the initialization of the multichannel ASR. Then, we trained the stream attention module of the multichannel ASR with 20 randomly selected channels per utterance. If the training utterances were from Libri-adhoc40-simu, the training condition is denoted as \textit{simu train}. If the training utterances were from Libri-adhoc40, the training condition is denoted as \textit{real train}.

We conducted the evaluation on the test data of Libri-adhoc40 in terms of word error rate (WER).

\subsection{Results}

\begin{figure}[t]
\centering
{
\begin{subfigure}{5cm}
\centering
\includegraphics[width=\linewidth]{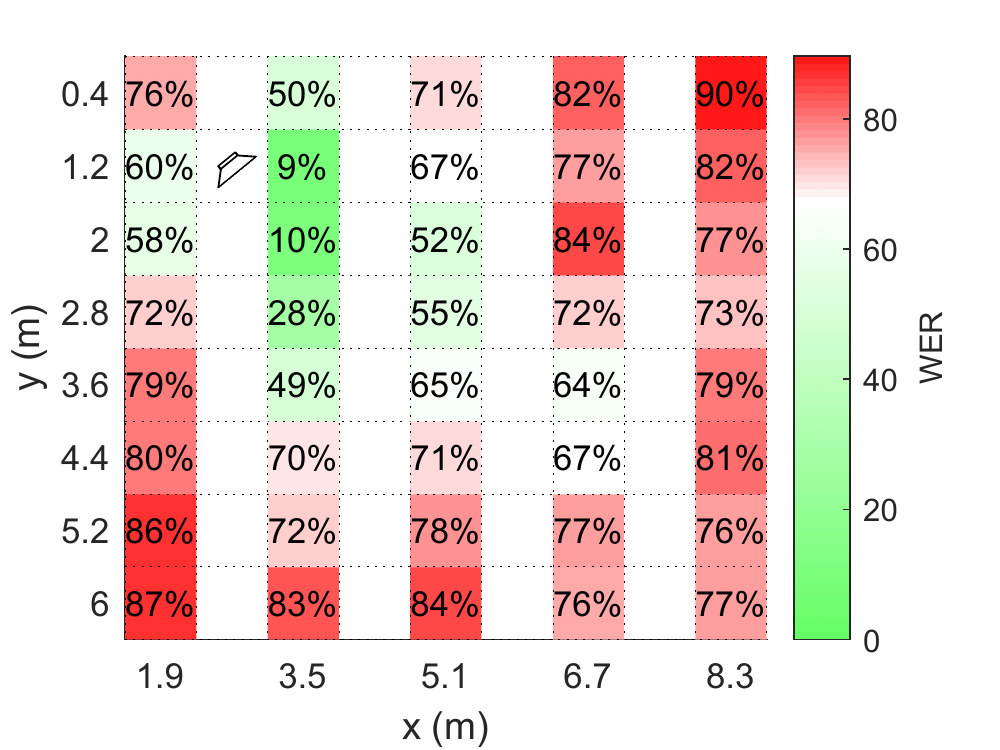}\label{fig:pos_wera}\vspace{-5mm}
\caption{Loudspeaker placed at pos 1}
\end{subfigure}\vspace{2mm}%
}%
\\
{
\begin{subfigure}{5cm}
\centering
\includegraphics[width=\linewidth]{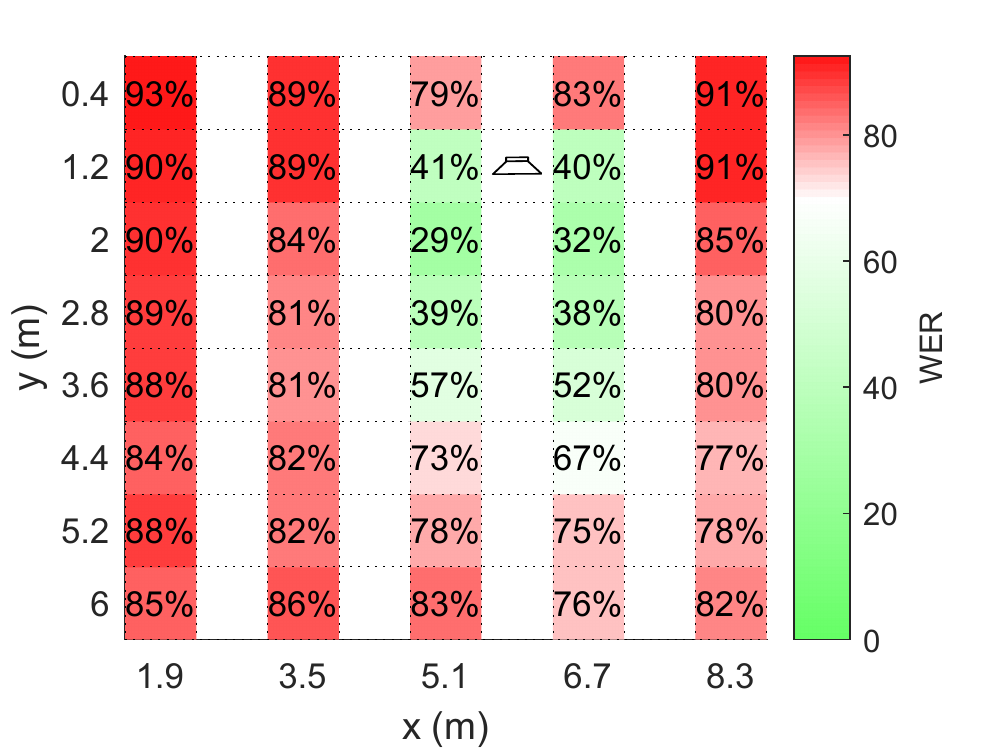}\label{fig:pos_werb}\vspace{-5mm}
\caption{Loudspeaker placed at pos 3}
\end{subfigure}
}
\caption{Visualization of the WER (\%) results of the single-channel conformer-based ASR system on the test data of Libri-adhoc40.}
\label{fig:pos_wer}
\end{figure}

Figure~\ref{fig:pos_wer} shows the average WER results of the single-channel conformer-based ASR system on each channel of the test set at two positions. From the figure, we see that (i) the average WER at the closest channel is 9\% in Figure~\ref{fig:pos_wer}(a) and 29\% in Figure~\ref{fig:pos_wer}(b); (ii) the WERs of the channels that the loudspeaker faces to in Figure~\ref{fig:pos_wer}(a) is significantly lower than those in Figure~\ref{fig:pos_wer}(b). The phenomena indicate that the performance was not only affected by the distance between the speaker and the microphone, but also affected by the orientation of the speaker.

Table~\ref{tab:wer_res} lists the comparison results when the ad-hoc microphone array contains 10, 25, and 40 channels respectively. From the table, we can see that the models, no matter trained on simulated data or semi-real data, can be used on the semi-real test data of the proposed Libri-adhoc40. The systems {in the \textit{real train} condition} perform better than those {in the \textit{simu train} condition}. When there is no microphone in the orientation of the loudspeaker, such as `pos2' and `pos3', all methods behave poorly.

Besides the general phenomena, the results can also reflect the trend and difference of the models with different ad-hoc microphone arrays. Specifically, (i) as the number of channels increases, the performance of the ASR systems is gradually improved. For example, the WER of the Scaling Sparsemax {in the \textit{real train} condition} is reduced by {46.6\%} relatively when the channel number is increased from 10 to 40, which demonstrates the importance of increasing the number of the channels. {(ii)} When the channel number is 40, the Scaling Sparsemax achieves a relative WER reduction of {19.6\%} lower than the oracle one-best {in the \textit{real train} condition}, which demonstrates the importance of channel selection. (iii) Although the orientation of the loudspeaker affects the performance significantly, Scaling Sparsemax reduces the negative effect. For example, when the channel number is 40, the average WER of the oracle one-best at `pos2' and `pos3' is increased by about $66\%$ over that at `pos1' and `pos4', while the relative WER increase of Scaling Sparsemax {in the \textit{real train} condition} is only $54\%$, which demonstrates the merit of ad-hoc microphone arrays.

To summarize, the above phenomena demonstrate the effectiveness of Libri-adhoc40 as an evaluation benchmark.

\begin{table}[t]
	\caption{Comparison results (in WER (\%)) on the test set of {Libri-adhoc40}. The term `pos\#' means that the test data is a subset of Libri-adhoc40 test data where the loudspeaker was placed at \emph{pos \#} described in Figure~\ref{fig:rec_setting}(b).}
	\label{tab:wer_res}
	\centering
\scalebox{0.9}{
\begin{tabular}{ccccccc}
\hline
\multicolumn{1}{c|}{\textbf{Method}}                                                              & \multicolumn{1}{c|}{\textbf{\begin{tabular}[c]{@{}c@{}}Training\\ condition\end{tabular}}} & \multicolumn{1}{c|}{\textbf{Pos1}} & \multicolumn{1}{c|}{\textbf{Pos2}} & \multicolumn{1}{c|}{\textbf{Pos3}} & \multicolumn{1}{c|}{\textbf{Pos4}} & \textbf{AVG} \\ \hline
\multicolumn{7}{c}{10 channels}                                                                                                                                                                                                                                                                                    \\ \hline
\multicolumn{1}{c|}{Oracle}                                                                       & \multicolumn{1}{c|}{Librispeech}                                                   & \multicolumn{1}{c|}{32.5}          & \multicolumn{1}{c|}{46.2}          & \multicolumn{1}{c|}{43.6}          & \multicolumn{1}{c|}{39.4}          & 40.4          \\ \hline
\multicolumn{1}{c|}{\multirow{2}{*}{\begin{tabular}[c]{@{}c@{}}Scaling\\ Sparsemax\end{tabular}}} & \multicolumn{1}{c|}{\textit{simu train}}                                                    & \multicolumn{1}{c|}{28.6}          & \multicolumn{1}{c|}{43.5}          & \multicolumn{1}{c|}{36.3}          & \multicolumn{1}{c|}{35.8}          & 36.1          \\ \cline{2-7}
\multicolumn{1}{c|}{}                                                                             & \multicolumn{1}{c|}{\textit{real train}}                                                    & \multicolumn{1}{c|}{\textbf{25.7}} & \multicolumn{1}{c|}{\textbf{38.5}} & \multicolumn{1}{c|}{\textbf{33}}   & \multicolumn{1}{c|}{\textbf{31.7}} & \textbf{32.2} \\ \hline
\multicolumn{7}{c}{25 channels}                                                                                                                                                                                                                                                                                    \\ \hline
\multicolumn{1}{c|}{Oracle}                                                                       & \multicolumn{1}{c|}{Librispeech}                                                   & \multicolumn{1}{c|}{12.4}          & \multicolumn{1}{c|}{33}            & \multicolumn{1}{c|}{32.4}          & \multicolumn{1}{c|}{16.9}          & 23.6          \\ \hline
\multicolumn{1}{c|}{\multirow{2}{*}{\begin{tabular}[c]{@{}c@{}}Scaling\\ Sparsemax\end{tabular}}} & \multicolumn{1}{c|}{\textit{simu train}}                                                    & \multicolumn{1}{c|}{12.5}          & \multicolumn{1}{c|}{32.9}          & \multicolumn{1}{c|}{27.4}          & \multicolumn{1}{c|}{17.4}          & 22.5          \\ \cline{2-7}
\multicolumn{1}{c|}{}                                                                             & \multicolumn{1}{c|}{\textit{real train}}                                                    & \multicolumn{1}{c|}{\textbf{12.3}} & \multicolumn{1}{c|}{\textbf{27.9}} & \multicolumn{1}{c|}{\textbf{24.4}} & \multicolumn{1}{c|}{\textbf{16.5}} & \textbf{20.3} \\ \hline
\multicolumn{7}{c}{40 channels}                                                                                                                                                                                                                                                                                    \\ \hline
\multicolumn{1}{c|}{Oracle}                                                                       & \multicolumn{1}{c|}{Librispeech}                                                   & \multicolumn{1}{c|}{9.1}           & \multicolumn{1}{c|}{31.7}          & \multicolumn{1}{c|}{32.2}          & \multicolumn{1}{c|}{12.6}          & 21.4          \\ \hline
\multicolumn{1}{c|}{\multirow{2}{*}{\begin{tabular}[c]{@{}c@{}}Scaling\\ Sparsemax\end{tabular}}} & \multicolumn{1}{c|}{\textit{simu train}}                                                    & \multicolumn{1}{c|}{\textbf{8.7}}  & \multicolumn{1}{c|}{29.1}          & \multicolumn{1}{c|}{24.9}          & \multicolumn{1}{c|}{12.9}          & 18.9          \\ \cline{2-7}
\multicolumn{1}{c|}{}                                                                             & \multicolumn{1}{c|}{\textit{real train}}                                                    & \multicolumn{1}{c|}{9.4}           & \multicolumn{1}{c|}{\textbf{25.3}} & \multicolumn{1}{c|}{\textbf{21.8}} & \multicolumn{1}{c|}{\textbf{12.3}} & \textbf{17.2} \\ \hline
\end{tabular}}
\end{table}

\section{Conclusions and discussion}

This paper presents a semi-real dataset recorded by synchronized ad-hoc microphone arrays, named Libri-adhoc40. Its validness has been evaluated in the speech recognition task. It facilitates the study and development of speech processing algorithms based on ad-hoc microphone arrays.

The dataset can be used as a benchmark corpus of many speech processing tasks beyond speech recognition, including speech enhancement, dereverberation, given that the anechoic recordings are provided in Libri-adhoc40. It can also be used for speech separation by mixing the speech signals at different positions of the loudspeakers, since that the loudspeakers at different positions replay different speakers. As for speaker recognition, we may take the entire set of 331 speakers for evaluation. As for the research on the synchronization techniques of devices, we may also construct asychronized test environments by adding various interruptions to the channels.

\bibliographystyle{IEEEtran}

\bibliography{mybib}

\end{document}